\documentclass[aps,prb,twocolumn,longbibliography]{revtex4-1}
\usepackage{epsfig,amsopn}
\usepackage{graphicx}
\usepackage{color}
\usepackage{amsmath,amssymb}
\usepackage{enumerate}
\newcommand\bea{\begin{eqnarray}}
\newcommand\eea{\end{eqnarray}}
\newcommand\beq{\begin{equation}}
\newcommand\eeq{\end{equation}}

\def\nn{\nonumber}
\def\f{\frac}

\def\ga{\gamma}

\def\si{\sigma}

\def\De{\Delta}

\def\la{\langle}
\def\ra{\rangle}
\def\ua{\uparrow}
\def\da{\downarrow}

\def\th{\theta}

\begin{document}
\title{Josephson diode effect in junctions of superconductors with band asymmetric metals}
\author{ Abhiram Soori}  
\email{abhirams@uohyd.ac.in}
\affiliation{ School of Physics, University of Hyderabad, Prof. C. R. Rao Road, Gachibowli, Hyderabad-500046, India}
\begin{abstract}
At  interfaces connecting two superconductors (SCs) separated by a metallic layer, an electric current is induced when there is a disparity in the phases of the two superconductors. We elucidate this phenomenon based on the weights of the Andreev bound states associated with the states carrying current in forward and reverse directions.  Typically,  current phase relation (CPR) in Josephson junctions is an odd function. When time reversal and inversion symmetries are broken at the junction, CPR ceases to be an odd function and the system may exhibit Josephson diode effect. This phenomenon has been studied in spin orbit coupled systems under an external Zeeman field wherein the magnetochiral anisotropy is responsible for the Josephson diode effect. Recently introduced the band asymmetric metal (BAM) model presents a novel avenue, featuring an asymmetric band structure.  We investigate  DC Josephson effect  in SC-BAM-SC junctions and find that band asymmetry can lead to Josephson diode effect and anomalous Josephson effect. We explain the mechanism behind these effects  based on interference of plane wave modes within the Bogoliubov de-Genne formalism. We calculate diode effect coefficient for different values of  the parameters. 
\end{abstract}
\maketitle

\section{Introduction}
A metallic or  a thin insulating material connecting two superconductors can carry an equilibrium current when a difference in the phases of the two superconductors is maintained. This phenomenon is known as dc Josephson effect~\cite{josephson62}. Bound states emerging within the superconducting gap transport a Josephson current when subjected to a phase bias~\cite{furusaki99,sahu23}.   Most often,  current phase relation (CPR)  of a Josephson junction is an odd function and the Josephson current is proportional to $\sin{\phi}$, where $\phi$ is the superconducting phase difference. 
Superconductors in proximity with ferromagnetic materials open up the possibility of nontrivial  triplet correlations~\cite{grein2009,Eschrig2007,Eschrig2008,braude2007,buzdin2008} which leads to harmonics other than $\sin{\phi}$ such as $\cos{\phi}$ giving rise to nonzero Josephson current at zero phase bias known as anomalous Josephson effect. Also, junctions between $s$-wave superconductor and spin-triplet superconductor, and Josephson junctions with magnetic interface on topological insulators exhibit this effect~\cite{asano2003,tanaka2009}. Interestingly, the supercurrent in Josephson junctions  with magnetic elements can be used to manipulate the magnetic moments~\cite{kon2009}. 

When time reversal ($\mathcal{TR}$) and inversion (${\mathcal I}$) symmetries are broken at the junction, CPR can change in a way that the maximum and minimum values of the  Josephson current $J(\phi)$ as the phase bias is varied do not add up to zero. This phenomenon is known as Josephson diode effect. Under ${\mathcal{TR}}$ symmetry, the phase bias $\phi\to-\phi$, which implies that if the junction is ${\mathcal{TR}}$ symmetric, $J(-\phi)=-J(\phi)$. Also under ${\mathcal I}$, $\phi\to-\phi$, which implies that if the junction is invariant under ${\mathcal I}$, $J(-\phi)=-J(\phi)$. This explains the reason why the junction needs to break ${\mathcal{TR}}$ and ${\mathcal I}$ for the Josephson diode effect to appear. 
In the expression for the Josephson current $J(\phi)$, contribution from terms proportional to $\sin{\phi}$, $\cos{\phi}$ and $\sin{2\phi}$ with same order of magnitude is necessary to get the diode effect~\cite{tanaka2022}. 
The last few years have witnessed a flurry of activities in superconducting- and Josephson- diode effects~\cite{siva18,yasuda19,ando2020,ita2020,baum2022,wu2022,suri22,souto22,soori23aje,soori23njde,moodera,Trahms2023,costa23}. 
 One way in which  ${\mathcal I}$ and ${\mathcal{TR}}$ symmetries are broken to achieve diode effect is magnetochiral anisotropy, wherein a Zeeman field is applied on a spin orbit coupled material~\cite{baum22mca,costa23}.  In this method, the velocities of the left- and the right- movers at the Fermi energy have different magnitude. In a recent experiment~\cite{ando2020}, the authors reported superconducting diode effect rooted in magnetochiral anisotropy. In another experiment~\cite{wu2022}, Josephson diode effect in absence of a magnetic field was reported, hinting at mechanisms other than magnetochiral anisotropy. This was followed by an experiment~\cite{moodera}  that reported superconducting diode effect at magnetic fields as low as $1$~Oe.

 \begin{figure*}[htb]
 \includegraphics[width=14cm]{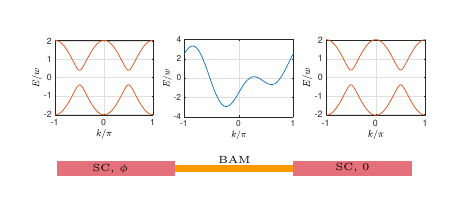}
 \caption{Schematic diagram of SC-BAM-SC junction, with the respective dispersions on the top. For BAM, only the electron band is shown.   }\label{fig:schem}
\end{figure*}

 Recently, a new model for a metallic system termed band asymmetric metal (BAM) was introduced~\cite{bam2023}. In such systems, the dispersion is asymmetric and the velocities of the left- and the right- movers have different magnitudes at any given energy. In this model, ${\mathcal{TR}}$ and ${\mathcal I}$ symmetries are broken by a complex next-nearest hopping amplitude on top of a nearest neighbor hopping in a one-dimensional lattice.  It has been found that the localization due to disorder is suppressed by asymmetry in the band. BAM when proximitised with a superconductor is known to exhibit superconducting diode effect~\cite{hosur23}. 

In this work, we study Josephson effect in  SC-BAM-SC  junctions, taking the model for BAM introduced in ref~\cite{bam2023}.  First, we comprehend Josephson effect in SC-metal-SC based on the currents carried by the Andreev states when the central metal has symmetric bands.  We then show that when the dispersion of the central metal is asymmetric, the Josephson diode effect shows up. The reason for the Josephson diode effect is rooted in the fact that the phases picked up by the Andreev  states between the two SCs carrying currents in the forward and the reverse directions are different in magnitude. Our work is closely related to the Josephson diode effect due to magnetochiral anisotropy since the mechanisms in the two are essentially the same in spirit.

\section{Calculation}
To enable a numerically  exact calculation, we describe the system on a finite one-dimensional lattice. The  system under study is depicted in Fig.~\ref{fig:schem}.  The Hamiltonian for SC-BAM-SC is: 
\bea 
H &=& H_L + H_{LB} + H_B + H_{BR} + H_R, \nn \\ 
  &&{\rm where} \nn \\
H_L &=& -w\sum_{n=1}^{L_S-1}[\Psi^{\dag}_{n+1}\tau_z\Psi_n + {\rm h.c.}]  \nn \\
 && -\mu_s\sum_{n=1}^{L_S}\Psi^{\dag}_n\tau_z\Psi_n + \De\sum_{n=1}^{L_S}\Psi^{\dag}_{n}(\cos{\phi}\tau_x+\sin{\phi}\tau_y)\Psi_n, \nn \\ 
H_{LB} &=& -w_J \Psi^{\dag}_{L_S+1}\tau_z\Psi_{L_S}, \nn \\ 
H_{B} &=& -w\sum_{n=L_S+1}^{L_{SB}-1}[\Psi^{\dag}_{n+1}\tau_z\Psi_n + {\rm h.c.}] -\mu_b\sum_{n=L_S+1}^{L_{SB}}\Psi^{\dag}_n\tau_z\Psi_n 
\nn \\ 
&& -w'\sum_{n=L_S+1}^{L_{SB}-2}[\Psi^{\dag}_{n+2}\tau_z e^{i\theta\tau_z}\Psi_n + {\rm h.c.}], \nn 
\eea
\bea
H_{BR} &=& -w_J \Psi^{\dag}_{L_{SB}+1}\tau_z\Psi_{L_{SB}}, \nn \\ 
H_{R} &=&-w\sum_{n=L_{SB}+1}^{L_{SBS}-1}[\Psi^{\dag}_{n+1}\tau_z\Psi_n + {\rm h.c.}]   \nn \\
 && -\mu_s\sum_{n=L_{SB}+1}^{L_{SBS}}\Psi^{\dag}_n\tau_z\Psi_n + \De \sum_{n=L_{SB}+1}^{L_{SBS}}\Psi^{\dag}_{n}\tau_x\Psi_n  , \nn \\ \label{eq:ham}
\eea 
where  $L_{SB}=L_S+L_B$, $L_{SBS}=2L_S+L_B$, the superconductors on the left and the right have $L_S$ sites, band asymmetric metal at the center has $L_B$ sites, $\Psi_n=[c_{\ua,n},  ~c_{\da,n}, ~-c^{\dag}_{\da,n}, ~c^{\dag}_{\ua,n}]^{T}$, $c_{\si,n}$ is annihilation operator for an electron with spin $\si$ at site $n$, $\tau_{x,y,z}$ are Pauli spin matrices acting on the particle-hole sector and $w$ is the nearest-neighbor hopping amplitude in the superconductors and the band asymmetric metal.  $w'e^{i\th}$ is the next-nearest-neighbor hopping amplitude in band asymmetric metal, $\mu_s$ ($\mu_b$) is the chemical potential in superconductor (band asymmetric metal) and $w_J$ is the hopping amplitude that connects the superconductors to band asymmetric metal. 
 
Bound states within the superconducting gap carry Josephson current at the junction between two semi-infinite superconductors~\cite{furusaki99}. Since we have taken the superconductors to be finite, we need to calculate the Josephson current by taking into account the current contributions from all the occupied  states~\cite{soori20pump}. The charge current is conserved in the band asymmetric metal region and hence the charge current operator can be written as $\hat J = -iew_J(\Psi^{\dag}_{L_S+1}\Psi_{L_S}-{\rm h.c.})/\hbar$. The eigenstates and eigenenergies of the Hamiltonian $(|u_j\ra,E_j)$'s can be calculated numerically exactly for each $\phi$. The Josephson current is then given by 
 \bea 
 J(\phi) &=& \sum_{j,~E_j<0} \la u_j|\hat J|u_j\ra 
 \eea  
 The diode effect coefficient is defined as $\ga=2(J_{max}+J_{min})/(J_{max}-J_{min})$, where  $J_{max}$ ($J_{min}$) is the maximum (minimum) value of the Josephson current in CPR.

 The dispersion of the central metallic region is $E=\mp[2w\cos{k}+2w'\cos{(\pm2k+\th})+\mu_b]$ for electron/hole bands. The bands are symmetric  either when $w'=0$ or when $\th=0, \pi$.  The bands are asymmetric when $\th\neq 0,\pi$ and $w'\neq 0$. In Fig.~\ref{fig:schem}, the dispersion for the electron band of BAM is shown in the central panel on top for $w'=0.8w, \theta=0.6\pi, \mu_b=0$.

\begin{figure}[htb]
 \includegraphics[width=6cm]{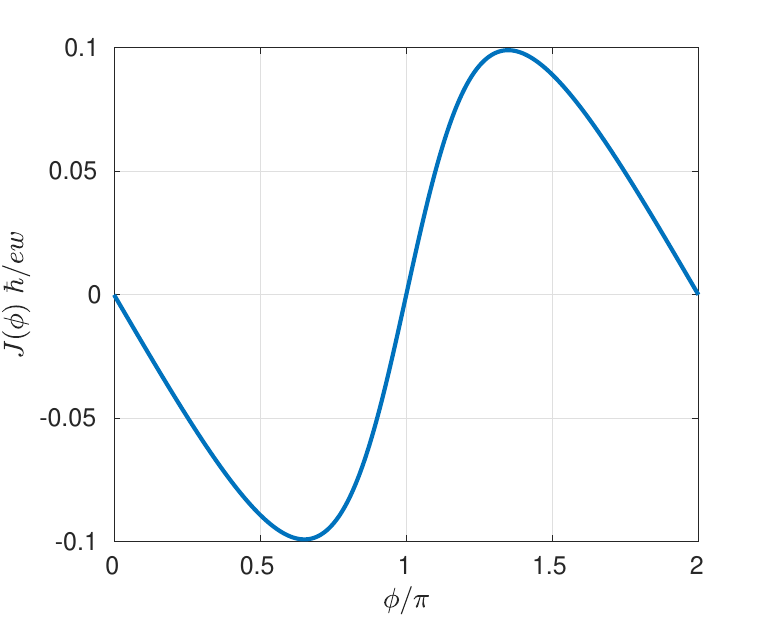}
 \caption{CPR of SC-metal-SC junction. The band of the central metal is symmetric. The current $J_{PW}(\phi)$ calculated from eq.~\eqref{eq:jpw} matches well with $J(\phi)$ and the two curves fall on top of each other. Parameters: $\mu_s=-w$, $\De=0.4w$, $L_S=L_B=10$, $w_J=0.9w$, $\mu_b=0$ and $w'=0$.}\label{fig:cprjj}
\end{figure}

\begin{figure*}[htb]
 \includegraphics[width=9cm]{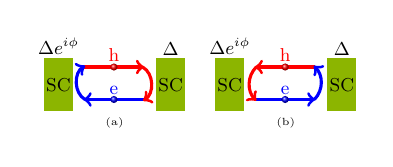}
 \includegraphics[width=4.2cm]{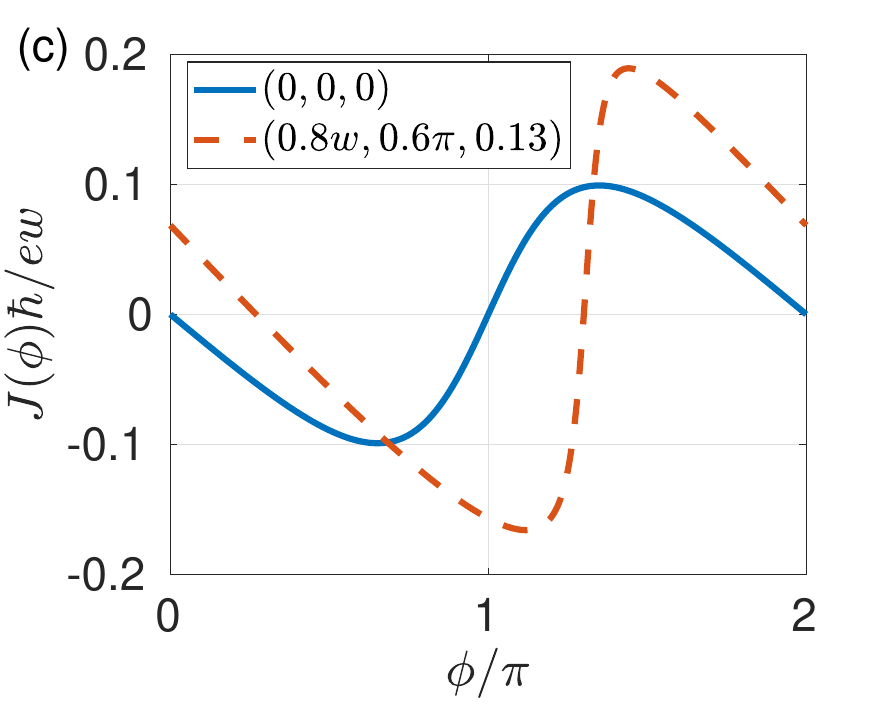}
 \includegraphics[width=4.2cm]{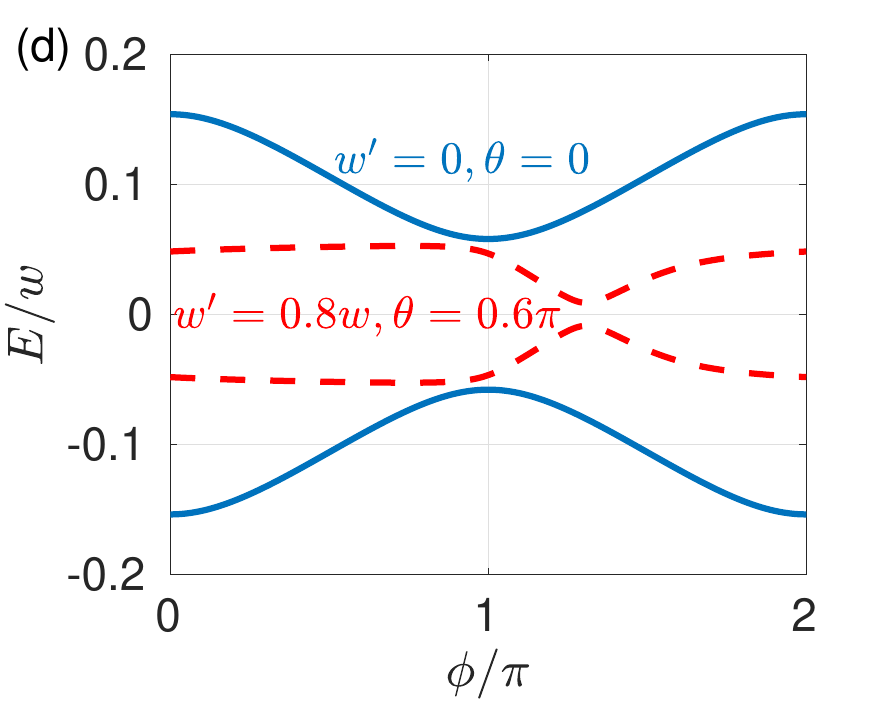}
  \includegraphics[width=19.2cm]{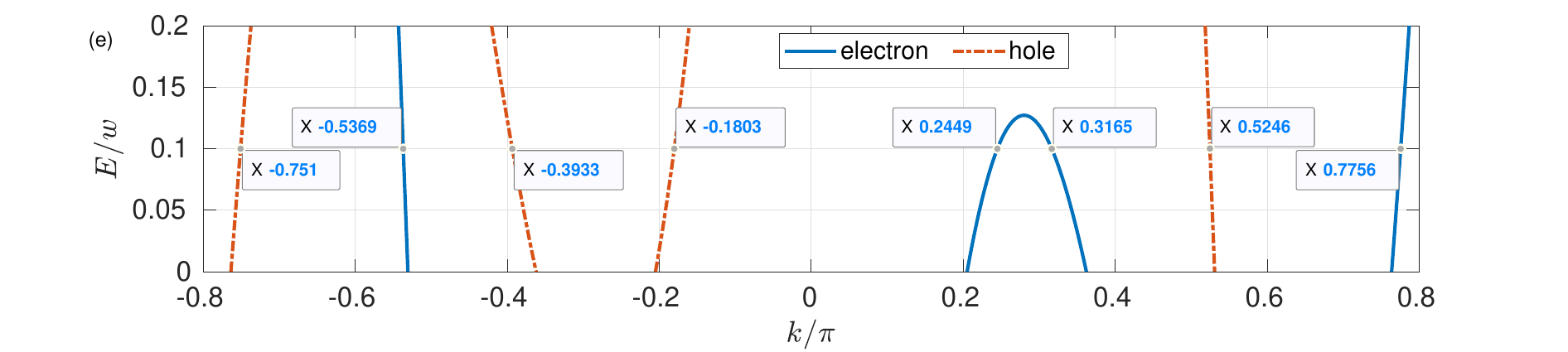}
 \caption{ (a \& b) Schematic of the two processes that carry current from one superconductor to the other under a phase bias $\phi$. In each of the diagrams above, the currents carried by the electron and hole are equal. But the currents carried by the processes in (a) and (b) are different in magnitude under a phase bias, leading to a net Josephson current. (c) CPR for the system described by the Hamiltonian in eq.~\eqref{eq:ham}. The two curves are for different values of $(w',\theta)$. The values of $(w',\theta,\gamma)$ are shown for each curve in the legend. When the band of central metal is asymmetric ($w'=0.8w, \phi=0.6\pi$), the Josephson diode effect shows up. (d) Energy of the states closest to zero energy versus the phase bias $\phi$. Asymmetry shows up  in this curve  when the band of the central metal is asymmetric. (e) The dispersions for the central band asymmetric metal zoomed in. The data points at energy $0.1w$ are chosen for the analysis. $x$-coordinates of all the points at energy $E=0.1w$ are shown. Parameters: $\mu_s=-w$, $\De=0.4w$, $L_S=L_B=10$, $w_J=0.9w$, $\mu_b=0$. }\label{fig:cpr-asym}
\end{figure*}

 \section{Josephson effect with band symmetric normal metal}

 Here, we elucidate on how the Josephson effect manifests in the system in absence of band asymmetry by choosing $w'=0$. CPR is plotted in Fig.~\ref{fig:cprjj} for the choice of parameters: $\mu_s=-w$, $\De=0.4w$, $L_S=L_B=10$, $w_J=0.9w$, $\mu_b=0$. The reason behind `why there is a nonzero supercurrent between SCs under phase bias' can be understood by the following argument. Let us consider an Andreev bound state at energy $E_l$. The electron and hole plane wave states with wave numbers $\pm k_{e,l}=\pm \cos^{-1}[-(\mu_b+E_l)/2w]$ and $\pm k_{h,l}=\pm\cos^{-1}[-(\mu_b-E_l)/2w]$ make up the Andreev bound state at energy $E_l$ in the metallic region. Let the wave function corresponding to this Andreev bound state be $|\psi_l\ra=\sum_{j=1}^{L_{SBS}}\sum_{\si}(\psi_{e,j,\si}|e,j,\si\ra+\psi_{h,j,\si}|h,j,\si\ra)$, where $|e,j,\si\ra$ ($|h,j,\si\ra$) corresponds to an electron (hole) of spin $\si$ at site $j$. The wave function corresponding to a right (left) moving electron in the central metal is $\sum_{j=L_S+1}^{L_{SB}}e^{ik_{e,l}}|e,j,\si\ra$ ($\sum_{j=L_S+1}^{L_{SB}}e^{-ik_{e,l}}|e,j,\si\ra$). The overlap of the state $|\psi_l\ra$ with a right moving (left moving) $\si$-spin electron plane wave in the central metal is $O_{l,\si,e,R} = \sum_{j=L_S+1}^{L_{SB}}e^{-ik_{e,l}j}\psi_{e,j,\si}/L_B$ ($O_{l,\si,e,L}=\sum_{j=L_S+1}^{L_{SB}}e^{ik_{e,l}j}\psi_{e,j,\si}/L_B$). These two overlaps have different magnitudes. This means that  the left moving electrons and the right moving electrons carry currents of different magnitudes. 
Further, the overlap of left moving (right moving) $\si$-spin  hole
$O_{l,\si,h,L}=\sum_{j=L_S+1}^{L_{SB}}e^{-ik_{h,l}j}\psi_{h,j,\si}/L_B$ ($O_{l,\si,h,R}=\sum_{j=L_S+1}^{L_{SB}}e^{ik_{h,l}j}\psi_{h,j,\si}/L_B$) has the property that: $|O_{l,\si,e,R}|^2\sin{k_{e,l}}=|O_{l,\bar{\si},h,L}|^2\sin{k_{h,l}}$ and  $|O_{l,\si,e,L}|^2\sin{k_{e,l}}=|O_{l,\bar{\si},h,R}|^2\sin{k_{h,l}}$, where $\bar{\si}$ is the spin opposite to $\si$. This means that the right (left) moving electron and left (right) moving hole carry equal currents. However,  the currents carried in the forward- and the backward- directions are not the same in magnitude under a phase bias, resulting in a net Josephson current. The Andreev processes for currents in the reverse and the forward directions are shown in Fig.~\ref{fig:cpr-asym}(a,b). The current contribution from the plane wave modes in the metal can be written as 
\bea 
J_{PW}(\phi) &=& \f{2ew}{\hbar}\sum_{l, *} \sum_{\si} 
[(|O_{l,\si,e,R}|^2 -|O_{l,\si,e,L}|^2)\sin{k_{e,l}} \nn \\ && + (|O_{l,\si,h,L}|^2 -|O_{l,\si,h,R}|^2)\sin{k_{h,l}}], \label{eq:jpw}
\eea 
where $\sum_{l,*}$ means that the summation is done over the states which obey the condition $(-2w-\mu_b)<E_l<0$. For this range of energies, the  states have plane wave nature in the central metallic region. 
We numerically find that the values of the currents $J(\phi)$ and $J_{PW}(\phi)$ agree well. It is possible that there are  states outside this energy range, but their contribution to the current is significantly lower.

\section{Josephson effect with band asymmetry in the metallic region}
In this section, we study Josephson effect when the band asymmetry in the central metallic region between the superconductors is turned on  by choosing $w'=0.8, \theta=0.6\pi$. Other parameters are the same as in the previous section. We show CPR in  Fig.~\ref{fig:cpr-asym}(c), and   the dependence of the energy level closest to zero energy on the phase bias $\phi$ in Fig.~\ref{fig:cpr-asym}(d). 
We find that the Josephson diode effect and an anomalous current show up when the band asymmetry is turned on. 
Also, the $E$ versus $\phi$ curve becomes asymmetric when the band asymmetry is switched on, as can be seen in Fig.~\ref{fig:cpr-asym}(d).

 To gain insight into the mechanism by which the diode effect manifests, we analyze the electron and hole dispersions in the central BAM.  Specifically, we examine the phases acquired by the processes carrying currents in both the forward and the reverse directions. We zoom in the dispersions in the energy range $(0,0.2w)$ inside the superconducting gap [which is $(-0.4w,0.4w)$] in Fig.~\ref{fig:cpr-asym}(e). We select the energy $E=0.1w$ for the analysis. The wave numbers for the right moving electrons and left moving holes are $(0.7756, 0.2449)\pi$ and $(-0.3933, 0.5246)\pi$. This means that the total phase picked up by the plane wave in one round of back-and-forth traversal is $0.8892\pi L_b$ for the state carrying current in the forward direction. On the other hand, the wave numbers for the left moving electrons and right moving holes are $(-0.5369,0.3165)\pi$ and $(-0.1803,-0.751)\pi$ respectively. This implies that the total phase picked up by the plane wave in one back-and-forth travel is $0.7109\pi L_b$. It is evident that the phases picked up by the states carrying currents in the forward and reverse directions are not the same in magnitude. The phases picked up determine the energy of the Andreev states.  Hence, the currents in the forward and the backward directions under opposite phase biases are not the same in magnitude, leading to diode effect. Also, under zero phase bias, current in one direction is favored over the other, resulting in an anomalous current. For $\th=\pi/2$ (along with $\mu_b=0$), the phases picked up for current carrying states in the forward and reverse directions are equal in magnitude. This nullifies the diode effect for this case, even though the band of the central metallic region is asymmetric. Hence, the band asymmetry of the central metallic region  is a necessary condition for Josephson diode effect rather than a sufficient condition. 

\begin{figure}[htb]
 \includegraphics[width=6.2cm]{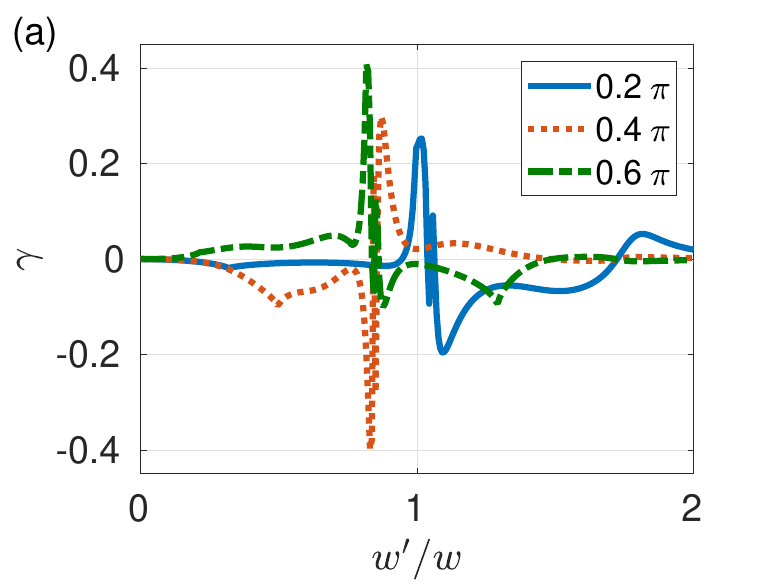}
 \includegraphics[width=6.2cm]{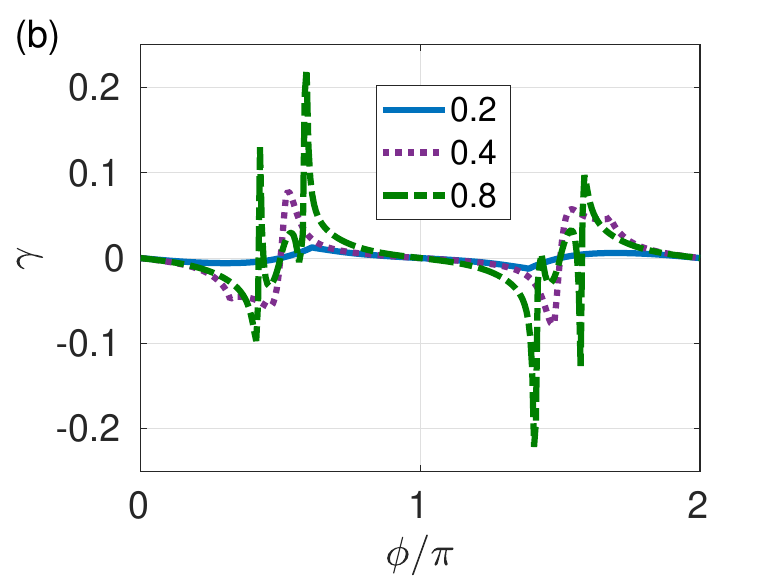}
 \caption{(a) Dependence of the diode effect coefficient  $\gamma$ on $w'$ - a parameter responsible for band asymmetry of the metal for different values of $\th$ indicated in the legend. (b) Dependence of $\ga$ on $\th$ - another parameter that dictates the band asymmetry of the metal for different values of $w'/w$ indicated in the legend. Other parameters are same as in Fig.~\ref{fig:cpr-asym}.}\label{fig:gamma}
\end{figure}

In Fig.~\ref{fig:gamma}, we show the dependence of diode effect coefficient on the parameters $w'$ and $\th$ of the band asymmetric metal. We find that the diode effect coefficient is not a monotonic function of $w'$ at a fixed $\th$ from Fig.~\ref{fig:gamma}(a). We also find that for $\th=\pi/2, 3\pi/2$ and $\mu_b=0$, the diode effect coefficient is zero. This can be explained by an argument similar to that for diode effect. The phases picked up by  the plane wave modes carrying currents in the forward- and  backward- directions have equal magnitudes for this case. 

\section{Discussion and Conclusion}
We have explained the DC Josephson effect in superconductor-metal-superconductor by an argument based on populations of the states carrying currents in the forward and reverse directions. When the band of the metal is asymmetric, CPR ceases to be an odd function. Further, an anomalous current and Josephson diode effect appear. However, for some special choice of parameters that make the band asymmetric, the diode effect and anomalous current vanish even though the band of the central metal is asymmetric. Hence, band asymmetry is a necessary condition for Josephson diode effect, but not a sufficient condition. We explain the diode effect by showing that the phases picked up by the states carrying current in the forward and reverse directions are different. We find that the anomalous Josephson effect accompanies the Josephson diode effect.  The model employed for BAM in this work~\cite{bam2023} is microscopic, distinguishing itself from the approach in ref.~\cite{fu22diode}, where the band asymmetry term is presumed in the Hamiltonian, leading to the manifestation of the Josephson diode effect. In systems with spin orbit coupling and magnetization, it is also possible that anomalous Josephson effect appears without diode effect when ${\mathcal{TR}}$ and ${\mathcal I}$ are broken~\cite{camp2015,Shukrinov2022}. This happens when the directions of spin orbit coupling and magnetization are perpendicular. Inversion in this setup is broken by barrier strengths at the two interfaces~\cite{camp2015}. The bands in the spin-orbit coupled region are not asymmetric in this case. Also, to  CPR, the harmonics that contribute are just $\sin{\phi}$ and $\cos{\phi}$. If the $\sin{2\phi}$-harmonic is also present, the CPR would exhibit Josephson diode effect.    But in BAM, both the effects show up together. Since localization by disorder is suppressed in BAM~\cite{bam2023}, our results are expected to survive weak disorder in BAM~\cite{tanaka2022}. A realistic system that mimics band asymmetry is a spin orbit coupled system with Zeeman field being parallel to the spin orbit coupling~\cite{bam2023}. A two dimensional model with this in spirit has been shown to exhibit anomalous Josephson effect and diode effect~\cite{yoko2014}. But their model considers impurities in the metal region and the explanation of  the two effects is rooted in scattering across the impurity. In contrast, we find the appearance of the two effects in an impurity free system and explain it with scattering phases acquired by the electrons and holes. Hence our work elucidates the mechanism of Josephson diode effect in spin orbit coupled systems with a Zeeman field~\cite{baum22mca,costa23}.  We hope that our work will aid researchers in the advancement of superconducting diode development. 

\acknowledgements 
AS thanks Subroto Mukerjee for useful discussions and for bringing Ref.~\cite{bam2023} to our notice.  AS thanks  SERB Core Research Grant (CRG/2022/004311) and  University of Hyderabad Institute of Eminence PDF for financial support. 
\bibliography{ref_bamjj}

\end{document}